\documentclass[prl,reprint,amsmath,amssymb,aps,superscriptaddress]{revtex4-1}
\pdfoutput=1
\usepackage{geometry,enumerate,amsmath,amssymb}
\usepackage{verbatim}
\usepackage{fullpage}
\usepackage{graphicx}
\usepackage{dcolumn}
\usepackage{bm}
\newcommand{\be}{\begin{equation}}
\newcommand{\ee}{\end{equation}}
\newcommand{\bea}{\begin{eqnarray}}
\newcommand{\eea}{\end{eqnarray}}
\newcommand{\bnabla}{\mbox{\boldmath $\nabla$}}
\newcommand{\xdiameter}{50} 
\newcommand{\xthickness}{1} 
\newcommand{\xdisc}{14} 
\newcommand{\xring}{28} 
\newcommand{\xlambda}{2.1}

\begin{document}
\title{Threaded rings that swim in excitable media}
\author{Antonio Cincotti}
\affiliation{Department of Mathematical Sciences, Durham University, Durham DH1 3LE, U.K.}
\affiliation{Department of Chemistry, Durham University, South Road, Durham, DH1 3LE, U.K}
\author{Fabian Maucher}
\affiliation{Department of Physics and Astronomy, Aarhus University, Ny Munkegade 120, DK 8000 Aarhus, Denmark}
\affiliation{Department of Mathematical Sciences, Durham University, Durham DH1 3LE, U.K.}
\affiliation{Joint Quantum Centre (JQC) Durham-Newcastle, Department of Physics, Durham University, Durham DH1 3LE, United Kingdom}
\author{David Evans}
\affiliation{Department of Mathematical Sciences, Durham University, Durham DH1 3LE, U.K.}
\author{Brette M. Chapin}
\affiliation{Department of Mathematical Sciences, Durham University, Durham DH1 3LE, U.K.}
\affiliation{Department of Chemistry, Durham University, South Road, Durham, DH1 3LE, U.K}
\author{Kate Horner}
\affiliation{Department of Mathematical Sciences, Durham University, Durham DH1 3LE, U.K.}
\affiliation{Department of Chemistry, Durham University, South Road, Durham, DH1 3LE, U.K}
\author{Elizabeth Bromley}
\affiliation{Department of Physics, Durham University, Durham DH1 3LE, U.K.}
\author{Andrew Lobb}
\affiliation{Department of Mathematical Sciences, Durham University, Durham DH1 3LE, U.K.}
\author{Jonathan W. Steed}
\affiliation{Department of Chemistry, Durham University, South Road, Durham, DH1 3LE, U.K}
\author{Paul Sutcliffe}
\affiliation{Department of Mathematical Sciences, Durham University, Durham DH1 3LE, U.K.}
\date{October 2019}

\begin{abstract}
Cardiac tissue and the Belousov-Zhabotinsky reaction provide two notable examples of excitable media that support scroll waves, in which a filament core is the source of spiral waves of excitation. Here we consider a novel topological configuration in which a closed filament loop, known as a scroll ring, is threaded by a pair of counter-rotating filaments that are perpendicular to the plane of the ring and end on the boundary of a thin medium. We simulate the dynamics of this threaded ring (thring) in the photosensitive Belousov-Zhabotinsky excitable medium, using the modified Oregonator reaction-diffusion equations. These computations reveal that the threading topology induces an exotic motion in which the thring swims in the plane of the ring. We propose a light templating protocol to create a thring in the photosensitive Belousov-Zhabotinsky medium and provide experimental confirmation that this protocol indeed yields a thring.
 \end{abstract}
\maketitle

Excitable media that host spiral wave vortices are found in a variety of chemical, biological and physical systems \cite{Winfree}.
An important example is cardiac tissue, where spiral waves 
are believed to play a vital role in certain cardiac arrhythmias \cite{Wit}.
Experimental studies in this context are clearly difficult, but fortunately
the Belousov-Zhabotinsky (BZ) chemical reaction provides a similar excitable medium that is more amenable to experimental study and therefore allows a detailed investigation of the properties of spiral waves \cite{Epstein}.
The photosensitive variant of the BZ reaction is particularly accommodating, as it allows some optical control of excitation waves. Similar optical control has recently been demonstrated in cardiac tissue \cite{Burton} and optogenetic
defibrillation has been shown to terminate ventricular arryhthmia \cite{Bruegmann}.

In a three-dimensional medium, spiral wave vortices become extended vortex strings, known as scroll waves \cite{Win73}, with a linelike filament that organizes waves of excitation. A filament that forms a closed circular loop is known as a scroll ring, with the generic motion being translation along the axis of the ring, akin to the familiar motion of a smoke ring. Locomotion of this type is thwarted in a thin medium, where tight confinement in the direction parallel to the axis of the scroll ring traps it at the medium boundary. Recent experimental results \cite{ATE}
in a thin photosensitive BZ medium have shown that such confinement not only prevents the translational motion of the scroll ring but also significantly changes other aspects of its dynamics, yielding a stable ring radius in the confined medium in contrast to an expanding ring in the corresponding bulk system.
The experimental results are well-described by numerical simulation of the modified Oregonator reaction-diffusion equations that model the photosensitive version of the BZ medium \cite{ATE}. The dramatic effects of spatial confinement on scroll rings has also been demonstrated in other BZ media, where the results are again reproduced by reaction-diffusion equations and a kinematical model has been proposed that provides a quantitative description of the interaction of a scroll ring with a medium boundary \cite{TES}.
This influence of confinement on excitable media is important in other contexts, particularly in the human heart \cite{CF}, where the thickness is of the order of the spiral wavelength, as in the chemical experiments.

\begin{figure}
\includegraphics[width=\columnwidth]{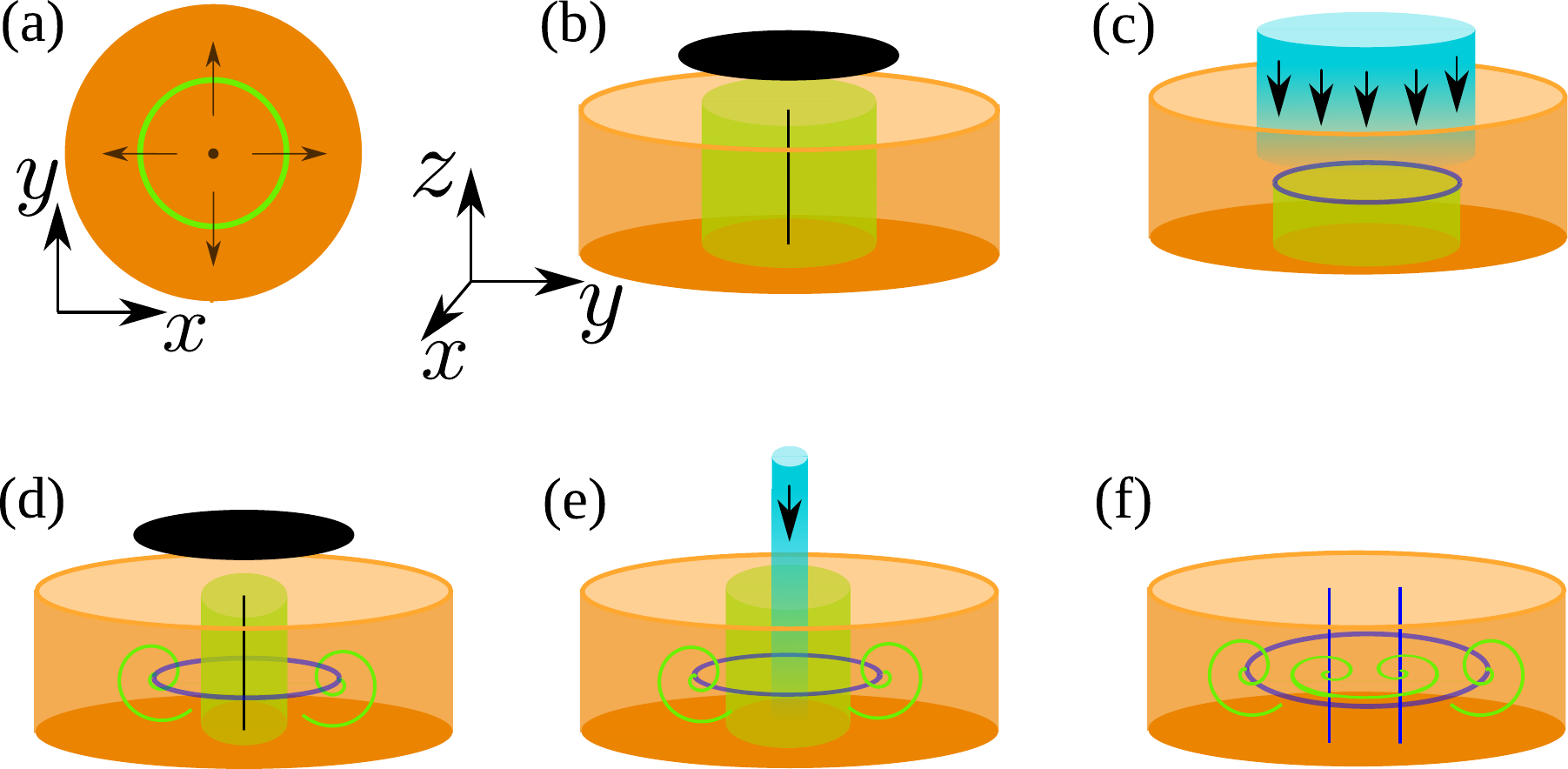}
\caption{
  Protocol for the initiation of a threaded ring: (a) and (b) A defect (black line) emits a cylindrical wave (green) from the centre. (c) Light is used to cut the top half of the emitted wave. (d) The circular end of the wave forms a scroll ring (blue) as another cylindrical wave emanates from the defect. (e) Light is used to cut a segment of the cylindrical wave to yield (f) a pair of vertical scroll wave filaments (blue) that thread the ring. 
  Note that the $z$ direction has been stretched to facilitate visualization.
  \label{fig:setup}
}
\end{figure}
\begin{figure*}
\includegraphics[width=2\columnwidth]{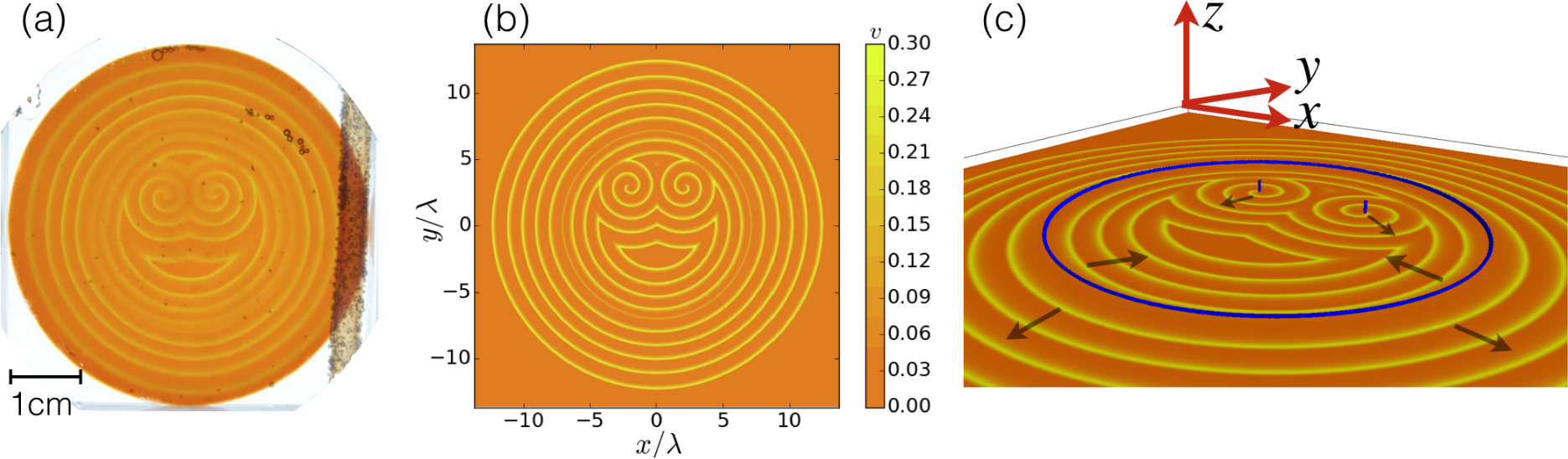}
\caption{
A threaded ring: (a) Experimental image. (b-c) Numerical simulation. (b) A heat map of the average value of $v$. (c) The scroll ring and threaded vertical scroll wave filaments are shown in blue, together with a heat map of $v$ on the bottom surface. The black arrows indicate the direction of motion of the excitation waves.
\label{fig:smile}
}
\end{figure*}

A scroll ring is the simplest example of a closed filament but knots and links provide more exotic examples, in which non-trivial topology profoundly influences the dynamics. To date, there are no experimental examples of knotted or linked filaments in any excitable medium. However, numerical simulations
  \cite{Winfree:Nature:1984,Winfree:SIAM:1990,Sutcliffe:PRE:2003,Sutcliffe:PRL:2016,Sutcliffe:PRE:2017,Sutcliffe:JPhysA:2018,Sutcliffe:Non:2019,BWA}
  provide predictions for their behaviour that are waiting to be confirmed by experiments, if methods can be engineered and implemented to create the required conditions and image the result.

  In this Letter we exploit confinement in a thin medium to provide the first experimental example of filaments with non-trivial topology in an excitable medium. We propose and implement a light templating protocol in a thin photosensitive BZ medium to create a scroll ring that is threaded by a pair of counter-rotating scroll waves which are perpendicular to the plane of the ring and end on the boundary of the thin medium. This rotaxane-like \cite{Bruns} arrangement of filaments is topologically non-trivial because the threading filaments cannot be unlinked from the circular filament of the ring whilst remaining unbroken and attached to opposite sides of the thin medium. For brevity, we shall refer to such a threaded ring using the portmanteau thring.

  The experimental results on the formation of a thring are reproduced by numerical simulations of the modified Oregonator reaction-diffusion equations that model the photosensitive version of the
 BZ medium \cite{Krug}. Our experimental apparatus requires that the thring is created with a transverse size of at least around ten spiral wavelengths. However, numerical simulations reveal an unexpected and novel phenomenon for smaller thrings, with a transverse size of around two spiral wavelengths. Namely, the combination of topology and confinement induces an exotic motion in which the thring swims in the plane of the ring, that is, in a direction perpendicular to the thin direction of the medium. The term thring is therefore an apt name, as the verb thring means to push ahead, as if in a throng.

Our protocol to create a thring is an extension of the approach used in \cite{ATE} to create a confined scroll ring and therefore we aim to recreate a similar experimental setup.
The photosensitive BZ medium incorporates a photosensitive ruthenium catalyst immobilized in a thin cylindrical silica hydrogel with diameter $\xdiameter\,{\rm mm}$ and thickness around $\xthickness\,{\rm mm}$.
The concentration of the ruthenium catalyst and the preparation of the gel followed the procedure described in \cite{Brand}.
The gel is positioned vertically in a glass chamber containing a catalyst-free BZ mixture with 
[NaBrO$_3$] = 0.2 M, [malonic acid] = 0.17 M, [H$_2$SO$_4$] = 0.39 M and [NaBr] = 0.09 M. Light is applied from one side of the gel using a video projector and images are captured on a camera positioned on the other side of the gel \cite{SM}. A bright light on a flexible head can be positioned between the camera and the gel, to locally increase the level of illumination and cut excitation waves as required. 
A circular plastic blocking disc, of diameter $\xdisc\,{\rm mm}$, can be attached to the side of the chamber facing the projector, to temporarily block the illumination inside this circular region. A speckle filter is placed between the video projector and the gel to yield a more homogeneous level of illumination at which the BZ medium is in an excitable regime. In the resting state the medium appears orange and a wave of excitation appears as light green, corresponding to the oxidized state of the catalyst.

Under a low light intensity the BZ reaction is oscillatory and spontaneously generates waves of excitation. The blocking disc can therefore be used to initiate spontaneous waves, however our protocol requires control over both the frequency of these wave emissions and over their precise point of generation. To achieve this control we introduce a novel experimental technique, where we shine a laser on a specific line through the gel for an extended period of time to photo-bleach a tiny region of the medium before it is placed in the chamber containing the BZ mixture. We refer to this photo-bleached region as a defect. Under low light intensity, waves will be emitted from the defect ahead of any other dark region. The frequency of the wave emission from the defect has been found to increase with the laser exposure time, allowing some control over the wave frequency. Once returned to ambient light, the defect ceases to produce waves and has no discernible influence on waves that pass through it.

Our protocol to create a thring is illustrated in Fig.\ref{fig:setup}. The blocking disc is centred over the defect to reduce the illumination and hence activate the defect, producing a cylindrical wave (Fig.\ref{fig:setup}ab). The bright light is used to cut the cylindrical wave (Fig.\ref{fig:setup}c), with the intensity and duration tuned to control the depth of the cut and subsequently form a scroll ring half way through the gel. The activity of the defect is tuned so that it produces a second cylindrical wave (Fig.\ref{fig:setup}d) at around the same time that the scroll wave is formed. Finally, the blocking disc is removed, to prevent further wave generation from the defect, and the bright light is used to cut a segment of the second cylindrical wave all the way through the gel (Fig.\ref{fig:setup}e), giving rise to a pair of oppositely handed scroll waves that thread the scroll ring and end on the boundaries of the medium (Fig.\ref{fig:setup}f).

In Fig.\ref{fig:smile}a we present the experimental realization of the above protocol with an image captured by the camera after several spiral wave rotation periods have elapsed since the creation of the thring. The pair of spiral scroll waves, with opposite orientations, are clearly visible near the centre of the gel. The filament of the scroll ring is the faintest of the circles, with a diameter of $\xring\,{\rm mm}$, and has the characteristic signature that the circular waves outside the scroll ring move away from the centre of the disc whereas the circular waves inside the scroll ring move towards the centre of the disc. The spiral wavelength is $\lambda=\xlambda\,{\rm mm}$, so the thickness of the gel is approximately $\frac{1}{2}\lambda$ and the diameter is $24\lambda.$

The experimental results are well-described by numerical simulations of the modified Oregonator reaction-diffusion equations \cite{Krug} that model the photosensitive version of the Belousov-Zhabotinsky medium, as in \cite{ATE}. In dimensionless form these equations are given by
\bea
\frac{\partial u}{\partial t}&=&\frac{1}{\varepsilon}\big(u(1-u)+w(\beta-u)\big)+\nabla^2 u, \
\frac{\partial v}{\partial t}=u-v, \nonumber \\
\frac{\partial w}{\partial t}&=&\frac{1}{\varepsilon'}\big(\Phi + \gamma v-w(\beta+u)\big)+\delta \nabla^2 w,
\label{pde}
\eea
where the variables $u,v,w$ are proportional to the concentrations of bromous acid, the oxidized form of the ruthenium catalyst, and bromide ions, respectively. We take the same parameter values as in \cite{ATE}, namely, 
$\varepsilon=0.125, \ \varepsilon'=0.00139, \ \beta=0.002, \ \gamma=1.16, \ \delta=1.12$ and  $\Phi=0.013.$
The modification from the original Oregonator model is the addition of the parameter $\Phi$ that is proportional to the light intensity, with the above value representing the light supplied by the projector. To model a region where light from the projector is blocked we set $\Phi=0$ in this region, and a region where the bright light illuminates the gel is modelled by setting $\Phi=0.04.$
Numerical solutions of (\ref{pde}) are computed using standard methods \cite{SM} with no-flux boundary conditions and support a spiral scroll wave with a wavelength $\lambda=18.6$ and a period $T=6.5.$ To compare with the experimental results we shall present all quantities in terms of these dimensions. In particular, the simulation region is a cuboid of size
$27\lambda\times 27\lambda\times \frac{1}{2}\lambda,$ and hence is comparable in size to the gel used in the experiment.

 \begin{figure}
\includegraphics[width=1.0\columnwidth]{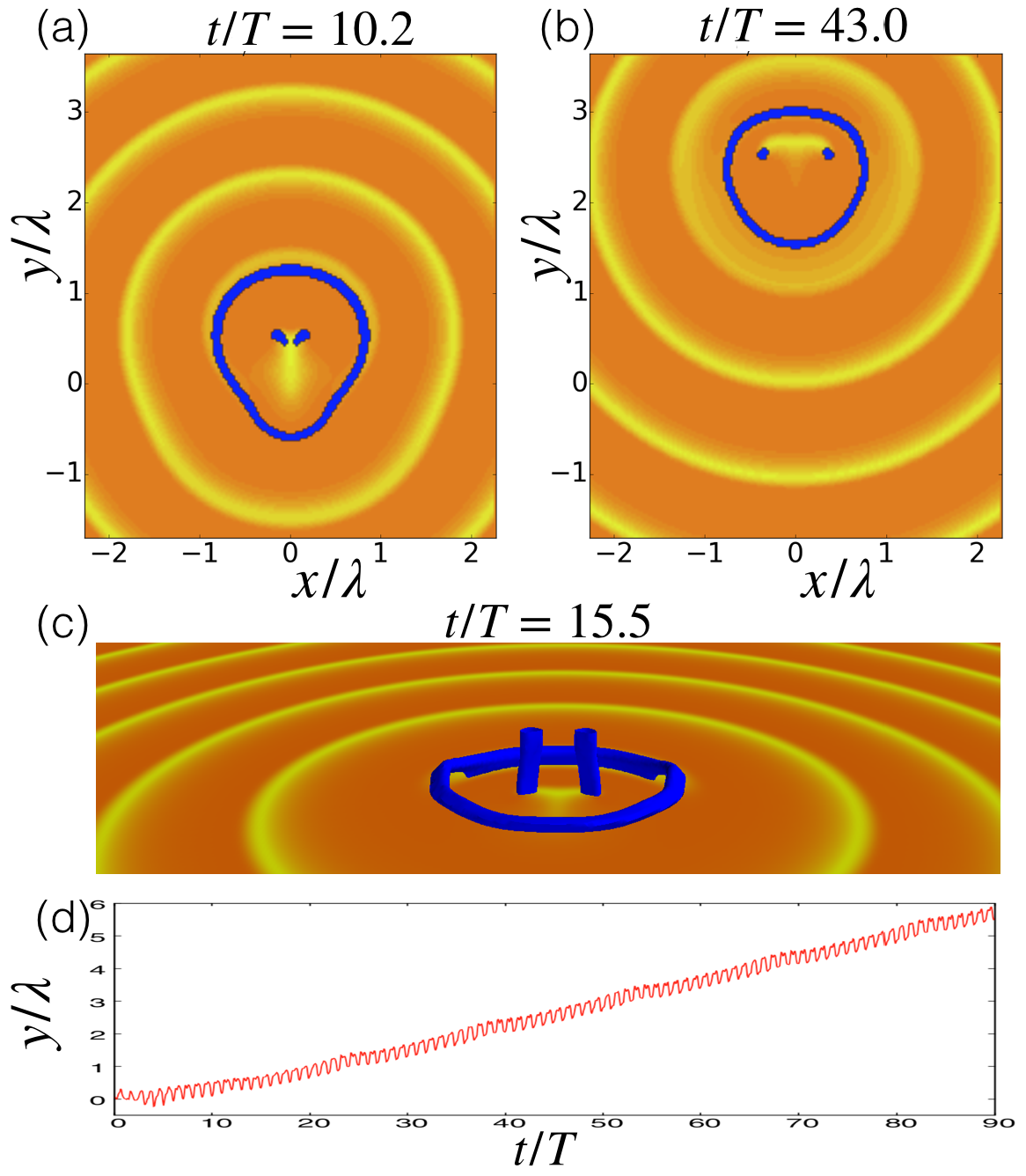}
\caption{A symmetric thring swimming along the $y$-axis. (a-b) The filaments (blue) and waves at increasing times. (c) A 3D view of the filaments. (d) The position along the $y$-axis of the thring as a function of time.
\label{fig:sym}
}
\end{figure}
In Fig.\ref{fig:smile}b we present the result of the numerical simulation of our protocol to create a thring, by displaying a heat map of the average value of $v$ along the $z$-direction. In the experimental image the colour change is proportional to the total integrated concentration of the oxidized catalyst through the thin direction of the gel, and hence to the average value of $v$ in the model. There is clearly a good agreement between the experimental result and the numerical simulation, with both displaying the smiling face image that is the signature of a thring of this size. The filaments can be visualized in the simulation by plotting the isosurface $|{\bm B}|=0.008,$ where ${\bm B} = \bnabla u \times \bnabla v$ is the vorticity that is highly localized on the filaments \cite{Win3}. In Fig.\ref{fig:smile}c the three filaments are shown in blue, together with a heat map of $v$ on the bottom surface of the medium.
In both the experiment and the simulation the dynamics of the thring is such that the threading filaments barely move while the filament ring slowly contracts \cite{SM}. In the simulation the ring retains its symmetry during the contraction but in the experiment the lack of perfect homogeneity in both the system and the initialization of the ring leads to an asymmetric contraction and eventually to a breaking of the filament ring due to a collision with the boundary of the gel, on a timescale of the order of 50 spiral wave periods (over an hour).  

In the numerical simulations we can study thrings which are smaller in size than those that our current experimental apparatus is capable of creating. The smallest possible thrings have a lateral size of around $2\lambda$ and demonstrate an unexpected novel form of locomotion.
In Fig.\ref{fig:sym}a-c, we present a thring at increasing times by displaying the filaments (blue), identified as regions where $|{\bm B}|\ge 0.008$, together with the excitation waves. These plots reveal that the thring moves along the $y$-axis, although the motion is far from rigid and the scroll ring filament takes on a variety of shapes,  with a motion that is reminiscent of swimming \cite{SM}.
The threading filaments are initially created  equidistant to the scroll ring. Choosing coordinates so that the centre of the scroll ring filament is at the origin and the threaded filaments are on the $x$-axis then the thring moves along the positive (negative) $y$-axis if the filament on the negative $x$-axis spirals clockwise (anti-clockwise). The $y$-coordinate of the centre of mass of the thring (calculated, as explained in
\cite{Sutcliffe:Non:2019}, by weighting positions by $|{\bm B}|^4$) is presented in Fig.\ref{fig:sym}d. Oscillations with the period $T$ are clearly seen on top of a constant swimming speed of $0.07\,\lambda/T,$ together with a secondary period $T_{\rm swim}\approx 15T$ that corresponds to a full stroke of the swimming action.

A qualitative understanding of the swimming mechanism can be obtained by appealing to the dynamics of the cores of a symmetric pair of oppositely handed spiral waves. By symmetry, the component of the velocity parallel to the line of symmetry, $v_\parallel$, is equal for each spiral core, whereas the component of the velocity perpendicular to the line of symmetry is $v_\perp$ for one core and $-v_\perp$ for the other, where $v_\perp>0$ corresponds to the cores moving away from each other. The magnitudes and signs of $v_\parallel$ and $v_\perp$ have a complicated dependence on the separation of the cores \cite{TES,Brand}. At the minimal separation at which the cores repel, $v_\parallel$ is similar to the speed of a swimming thring. However, because $v_\perp>0$ the cores move apart and this produces a drop in $v_\parallel$ by orders of magnitude and a reversal of its sign. 
 If the surrounding ring is removed from a thring then there is a short initial motion of the threading filaments as before, followed by a reversal of direction and a decay of the speed. In this case the centre of mass of the pair effectively comes to a halt, moving a total distance of less than $0.1\lambda$ in the total time displayed in Fig.\ref{fig:sym}d, in agreement with the above explanation for the swimming mechanism. The main effect of the ring in a small thring is to frustrate the repulsion of the cores, hence enabling a prolonged period of propulsion.

The size requirement for swimming behaviour follows from the above reasoning. The separation of the threading filaments must be less than $\lambda$, to have a significant parallel component to the velocity, and the distance from each threading filament to the ring must also be less than $\lambda$, to frustrate the repulsion. This yields a maximal size for a swimming thring of the order of $2\lambda.$ This is similar to the minimal size for a thring, because a pair of counter-rotating spiral waves created with a separation much less than $\lambda$ leads to their rapid mutual annihilation \cite{Ruiz,TES}.
To support swimming thrings the BZ medium must have a thickness of the order of $\lambda/2.$ The thickness cannot be less than this if it is to host a scroll ring and once the thickness is as large as $\lambda$ then the scroll ring loses its ability to tightly bind the threading filaments \cite{SM}.

\begin{figure}
\includegraphics[width=\columnwidth]{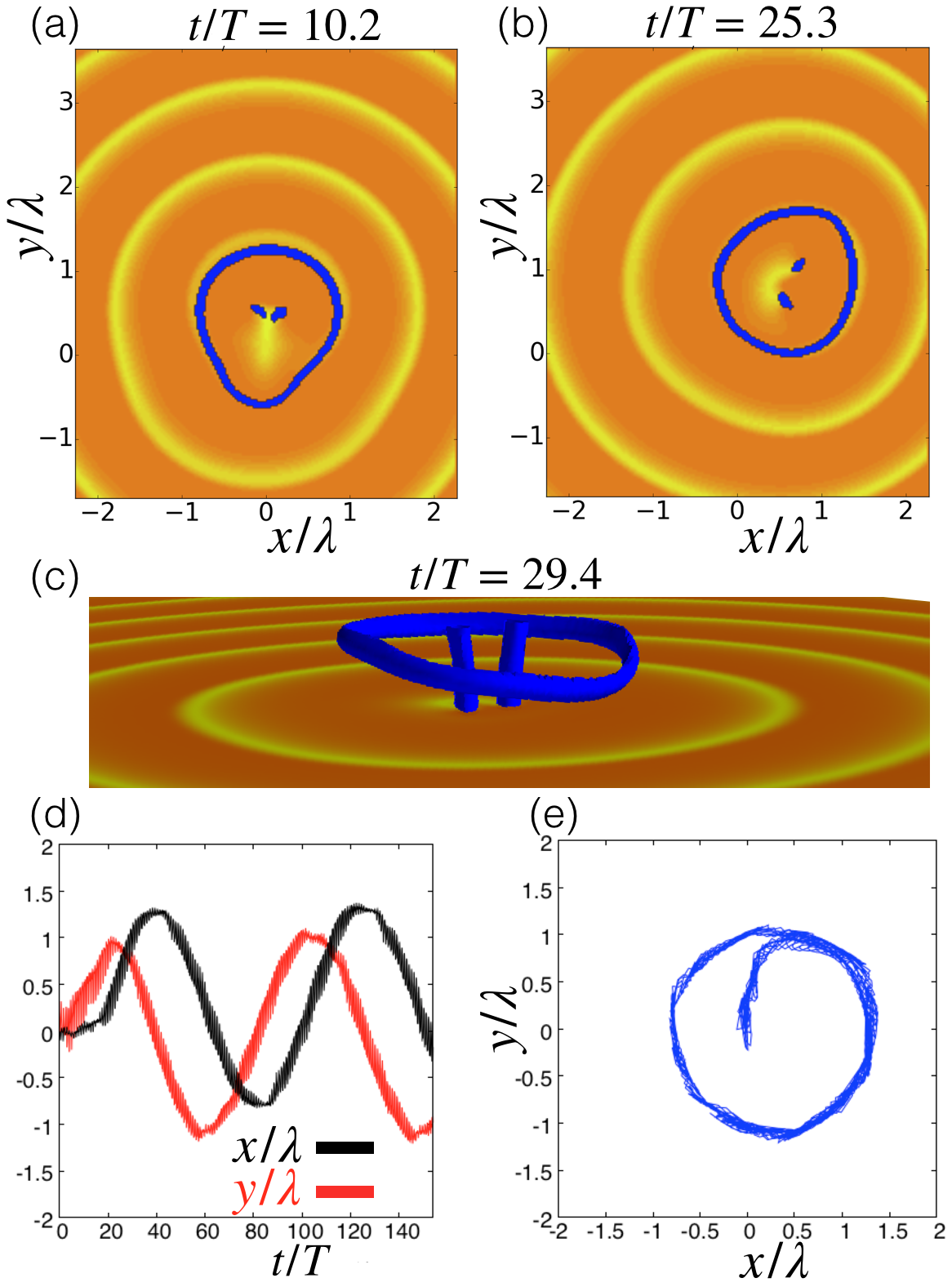}
\caption{An asymmetric thring swimming in a circle. (a-b) The filaments (blue) and waves at increasing times. (c) A 3D view of the filaments. (d) The $x$ (black) and $y$ (red) coordinates of the thring as a function of time. (e) The trajectory in the $(x,y)$-plane.
\label{fig:asym}
}
\end{figure}
It turns out that swimming in a fixed direction is an unstable locomotion for the thring and is a result of the symmetric initial conditions that created the threading filaments equidistant from the ring filament. Any small perturbation that breaks this symmetry and creates the clockwise spiral threading filament closer to the filament ring than its anti-clockwise partner yields a thring that swims in a circle in a clockwise direction. This is illustrated in Fig.\ref{fig:asym}, where the filaments and waves are displayed at two different times in  Fig.\ref{fig:asym}ab. The coordinates of the centre of mass of this asymmetric thring are plotted as a function of time in Fig.\ref{fig:asym}d and the circular trajectory is shown in Fig.\ref{fig:asym}e. There is an initial transitory phase, as the asymmetry develops, followed by circular motion with a diameter that is comparable to the size $2\lambda$ of the thring \cite{SM}. 
The time taken to swim a complete orbit is $T_{\rm orbit}\approx 86T,$ which is consistent with the swimming speed quoted above, $2\lambda\pi/T_{\rm orbit}\approx
0.07\,\lambda/T.$
The length of the transitory phase, where motion shifts from linear to circular, increases as the size of the initial asymmetry decreases, but the radius of the final circular trajectory is independent of the initial perturbation. As expected, if the anti-clockwise spiral is the one formed closest to the filament ring then the thring swims in a circle in an anti-clockwise direction.

In summary, we have introduced a light templating protocol to create a new topologically interesting arrangement of spiral wave filaments in a confined BZ excitable medium. We have experimentally realized a threaded ring (thring), as a proof of principle that these objects are experimentally accessible, and have demonstrated numerically that the interplay between threading topology and confinement induces a novel swimming motion for small thrings. To bring the experiment into the swimming regime would require an increase in the frequency of the defect or a decrease in the wave speed, so that the filament ring had a smaller radius at the time when the defect generates its second wave. This may be possible by making changes to the system, such as using different concentrations for the BZ mixture.

The experimental creation of filament knots in the BZ reaction remains an elusive goal for now. However, we believe that perfecting and extending our protocol  is an important step towards this realistic objective, using similar light templating protocols.
Swimming thrings also exist in the FitzHugh-Nagumo medium \cite{SM}, which provides a simple model of the electrical activity of cardiac tissue, and therefore they appear to be a new universal phenomenon in excitable media with potential implications across biology, physics and chemistry.

\section*{Acknowledgements}
This work is funded by the
Leverhulme Trust Research Programme Grant RP2013-K-009, SPOCK: Scientific Properties Of Complex Knots. F.M. acknowledges funding by the Danish National Research Foundation through a Niels Bohr Professorship to Thomas Pohl. 
The computations were performed on Hamilton, the Durham University HPC cluster.\\

\

\onecolumngrid
\renewcommand{\theequation}{S\arabic{equation}}
\renewcommand{\thefigure}{S\arabic{figure}}
\setcounter{figure}{0}
\renewcommand{\thesection}{\Roman{section}}
\renewcommand{\thesubsection}{\thesection.\arabic{subsection}}
\renewcommand{\thesubsubsection}{\thesubsection.\arabic{subsubsection}}
\newpage 

\section*{Supplemental Material}

\section{Additional Experimental Details}
The camera used to capture the images from the experiment was a Canon EOS 1300D 18 megapixel digital camera. The camera was positioned $15\,{\rm cm}$ from the gel and a picture taken every 10 seconds. Excitation waves are visible throughout the $1\,{\rm mm}$ thickness of the gel when illuminated using an EUG 5000 Lumens LCD LED 1080P Projector. The bright light used to cut the excitation waves was a Schott KL 1500 LCD $150\,{\rm W}$ halogen cold light source.
The laser used to photo-bleach the defect into the gel was a Thorlabs CPS450 collimated laser diode module, $450\,{\rm nm}$, $4.5\,{\rm mW}$, with an exposure time of $8$ hours.

With regard to the success rate of the experimental protocol, we were able to experimentally create 10 examples of thrings. The main problem to overcome is to create the scroll ring in the middle of the thin gel without a tilt angle, which requires a high level of homogeneity in both the reaction and the light source. Otherwise, we find that there is a quick breaking of the filament ring due to a collision with the boundary of the gel. This happens in about $50\%$ of the attempted initializations.  

\section{Numerical Methods}
Numerical solutions of the
modified Oregonator reaction-diffusion equations are computed using
an explicit fourth-order Runge-Kutta method with a timestep $dt=0.005$ and a 27 point stencil finite difference approximation for the Laplacian, with a lattice spacing $dx=0.5$ on a rectangular grid containing $1024\times 1024\times 20$ lattice points. No-flux (Neumann) boundary conditions are imposed at all boundaries of the medium.

\section{Thrings in a thicker medium}
\begin{figure}[h!]
\includegraphics[width=0.62\columnwidth]{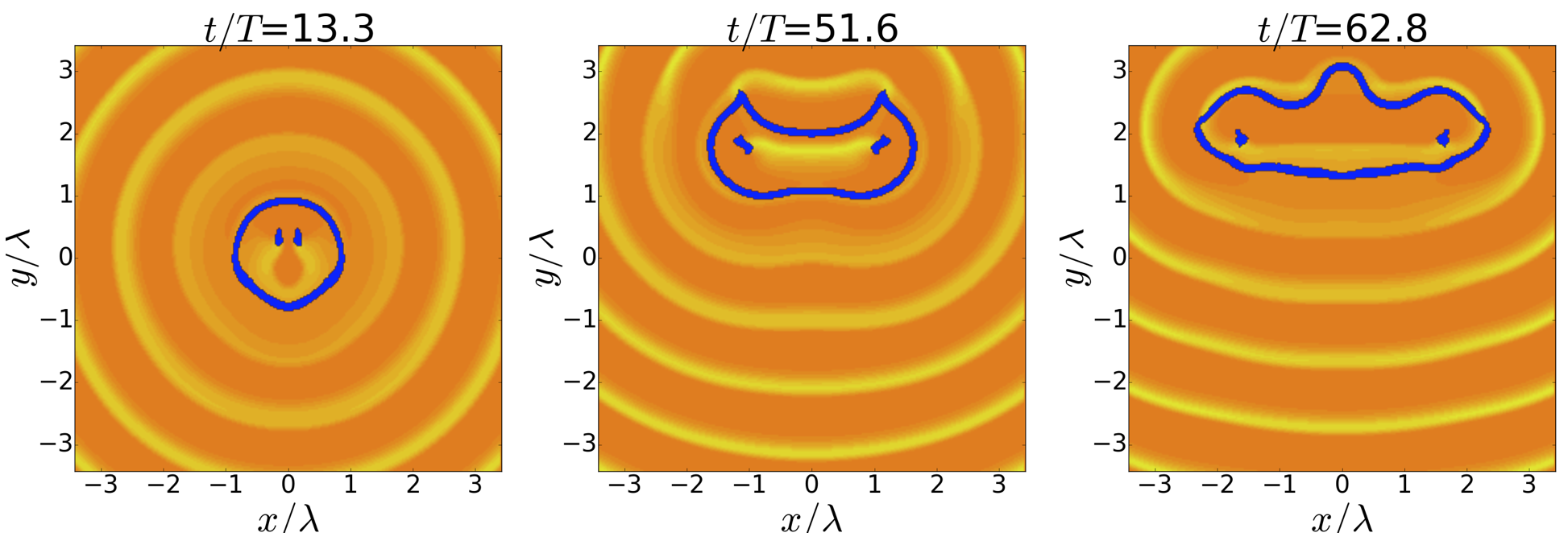}
\caption{A simulation that demonstrates the instability of a thring in a thicker medium. (a-c) The filaments (blue) and waves at increasing times.
  \label{fig:DOUBLE}
}
\end{figure}
Here we investigate the possibility of swimming thrings in a medium that is twice as thick as the previously studied  medium. Fig.\ref{fig:DOUBLE} displays the results of a simulation in a medium with thickness $\lambda.$ In this case the thring is unstable and the length of the scroll ring filament increases with time, as does the separation of the pair of threading filaments. The thring initially swims but the locomotion halts as the scroll ring is unable to tightly bind the threading filaments.

\section{Videos}
The experiment corresponding to Fig.2a is shown in the video {\em experiment.mp4}, where the time elaspsed between each frame is 10 seconds.
The simulations corresponding to Fig.2b,  Fig.3 and Fig.4 are shown in the videos
{\em simulation.mp4}, 
{\em swim1.mp4}, 
and {\em swim2.mp4}.

\section{Swimming Thrings in the FitzHugh-Nagumo model}
The FitzHugh-Nagumo equations provide a simple mathematical model of cardiac tissue as an excitable medium. The nonlinear reaction-diffusion partial differential equations are given by
\be
\frac{\partial u}{\partial t}=\frac{1}{\varepsilon}(u-\frac{1}{3}u^3-v)+\nabla^2 u,
\qquad
\frac{\partial v}{\partial t}=\varepsilon(u+\beta-\gamma v),
\ee
where $u$ represents the electric potential and
$v$ is the recovery variable.
The remaining variables are constant parameters that we fix to be
$\varepsilon=0.3, \ \beta=0.7, \ \gamma=0.5$, to give a spiral wavelength $\lambda=21.3$ and period $T=11.2.$

We compute numerical solutions of the FitzHugh-Nagumo equations using the same scheme used for the modified Oregonator model, as described in the numerical methods section. The simulation region is a cuboid of size $6\lambda\times 6\lambda\times \frac{1}{2}\lambda.$

\begin{figure}[h!]
\includegraphics[width=0.5\columnwidth]{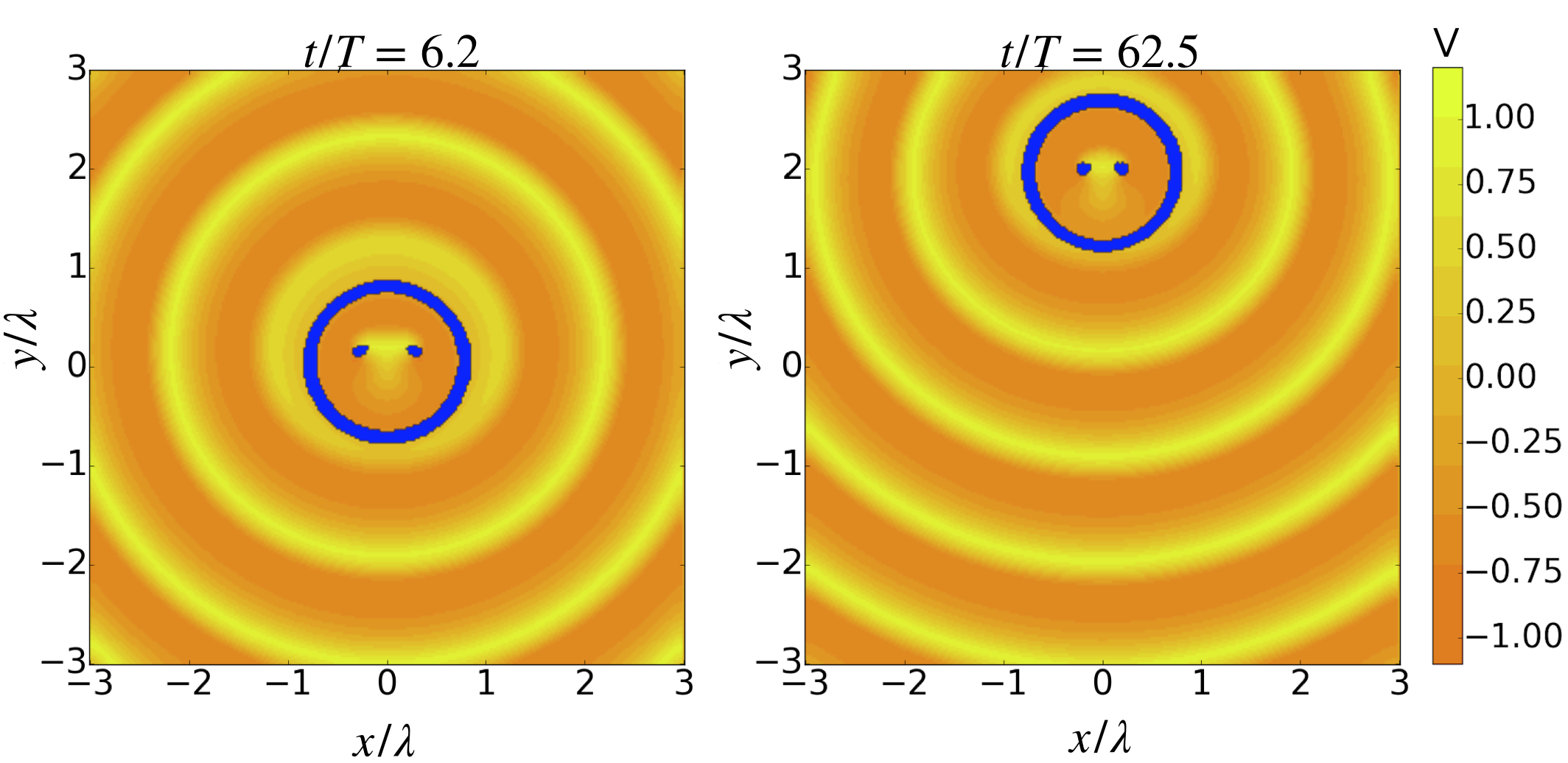}
\caption{A thring swimming along the $y$-axis in the FitzHugh-Nagumo model. (a-b) The filaments (blue) and waves at increasing times.
  \label{fig:FHN}
}
\end{figure}

In Fig.\ref{fig:FHN} we present a swimming thring at increasing times. The waves are again visualized by displaying a heat map of the average value of $v$ along the $z$-direction and the filaments are shown in blue as the regions where
$|{\bm B}|\ge 0.24.$ The fact that the FitzHugh-Nagumo model also hosts swimming thrings suggests that they are a universal phenomenon in excitable media.

\end{document}